\documentclass[aps,prl,preprint,groupedaddress]{revtex4-1}
\usepackage[dvips]{graphicx}
\begin{document}

\title{Growth of superconducting epitaxial films of sulfur substituted FeSe via pulsed laser deposition} 

\author{Fuyuki Nabeshima}
\email[]{nabeshima@maeda1.c.u-tokyo.ac.jp}
\affiliation{Department of Basic Science, the University of Tokyo, Meguro, Tokyo 153-8902, Japan}

\author{Tomoya Ishikawa}
\affiliation{Department of Basic Science, the University of Tokyo, Meguro, Tokyo 153-8902, Japan}

\author{Ken-ichi Oyanagi}
\affiliation{Department of Basic Science, the University of Tokyo, Meguro, Tokyo 153-8902, Japan}

\author{Masataka Kawai}
\affiliation{Department of Basic Science, the University of Tokyo, Meguro, Tokyo 153-8902, Japan}

\author{Atsutaka Maeda}
\affiliation{Department of Basic Science, the University of Tokyo, Meguro, Tokyo 153-8902, Japan}

\date{\today}

\begin{abstract}
We report the successful growth of epitaxial thin films of FeSe$_{1-x}$S$_x$ with $x \leq 0.43$ via pulsed laser deposition.
As S content increases, the nematic transition temperature, $T_{\mathrm s}$, decreases systematically and the superconducting transition temperature, $T_{\mathrm c}$, shows a gradual decrease even when $x$ exceeds the nematic end point (NEP), similar to bulk samples.
A new kink anomaly was observed in the $\rho$-$T$ curves for films with large $x$, which is likely due to a magnetic transition.
The obtained phase diagram of FeSe$_{1-x}$S$_x$ thin films is in contrast to that of FeSe$_{1-y}$Te$_y$ films, which shows a rapid increase of $T_{\mathrm c}$ at the NEP.
Our results demonstrate that the relation between the nematic order and the superconductivity is not universal in the FeSe system, suggesting that the nematic transition does not play a primary role in the superconductivity in these materials.
\end{abstract}


\maketitle

One of the current problems in the field of superconductivity is the role of electronic nematicity in superconductivity of iron based materials. 
An iron chalcogenide superconductor, FeSe\cite{Wu08}, which is one of the iron based superconductors (FeBSs) with the simplest crystal structure, shows a structural transition from tetragonal to orthorombic phase at 90 K\cite{PhysRevLett.103.057002}, below which an orbital ordered state was observed\cite{PhysRevB.90.121111,PhysRevLett.113.237001}.
This structural transition has an electronic origin, and thus is also called the nematic transition.
FeSe exhibits no long-range magnetic order at ambient pressure, while many other iron based superconductors show a magnetic transition at temperature very close to the nematic transition temperature\cite{NPhys.10.97}. 
Therefore, FeSe is considered to be one of  the most suitable materials for investigating the relation between the nematicity and the superconductivity in iron based superconductors.

Intense studies have been done on the chemical substitution of Se by isovalent S in FeSe\cite{PhysicaC.469.297,JPSJ.78.074712,JPCS.400.022125,PRB.91.165109,JPSJ.84.024713,PRB.92.121108,PRB.92.235113,PSS.254.1600153,PNAS.113.8139,PhysicaC.530.55,JLTP.185.467,arXiv:1611.07424,arXiv:1608.01044,PhysRevB.93.104502,PRB.93.224508,PRB.94.094506,PRL.117.157003,SuST.30.035017,NatCom.8.1143,JSNM.30.763,JJAP.56.100308,arXiv:1710.09892,arXiv:1710.02276,PRB.96.024511,PRB.96.020502,PRB.96.121103,arXiv:1802.03776,PNAS.115.1227}, in particular since a growth technique of FeSe single crystals by vapor transport was established\cite{PRB.87.180505}.
With increasing S content, the nematic transition temperature decreases monotonically, and the superconducting transition temperature, $T_{\mathrm c}$, slightly increases up to $\sim$ 11 K  and then starts to decrease.
While there is no significant change in $T_{\mathrm c}$ at the nematic end point (NEP), recent thermal measurement\cite{PNAS.115.1227} and scanning tunneling microscopy/spectroscopy measurement\cite{arXiv:1710.02276} has revealed that the superconducting gap shows an abrupt change at the NEP. 
These results suggest that the nematicity affects the superconductivity in FeSe to some extent. 

On the contrary, other isovalent substitution by Te results in a rather different behavior.
Although it is well-known that bulk samples of FeSe$_{1-y}$Te$_y$ are not available in the composition region of $0.1 \leq y \leq 0.4$ because of phase separation\cite{Fang08}, we have previously demonstrated the successful growth of single crystalline thin films with all $y$ values, including compositions inside the miscibility gap of bulk samples, via pulsed laser deposition\cite{yi15pnas,yiSciRep17}.
We have grown FeSe$_{1-y}$Te$_y$ films on two substrates, CaF$_2$ and LaAlO$_3$ (LAO), and the two phase diagrams obtained for the two different substrates show qualitatively the same behavior: the Te substitution decreases the nematic transition temperature, and $T_{\mathrm c}$ increases drastically just after the disappearance of the nematic transition\cite{yiSciRep17}. 
These observations in FeSe$_{1-y}$Te$_y$ films is in contrast to the case of FeSe$_{1-x}$S$_x$ bulk crystals, which does not show such a $T_{\mathrm c}$ enhancement, or rather $T_{\mathrm c}$ decreases at the NEP. 

There are two possibilities for the difference in the behavior of $T_{\mathrm c}$ between bulk FeSe$_{1-x}$S$_x$ and FeSe$_{1-y}$Te$_y$ thin films: difference between S and Te or between bulk and film. 
To judge which is true, it is important to grow FeSe$_{1-x}$S$_x$ thin films and to obtain their phase diagram.
For instance, FeSe$_{1-x}$S$_x$ thin films may exhibit a $T_{\mathrm c}$ jump at the NEP as well as the FeSe$_{1-y}$Te$_y$ films.
Another unsatisfactory feature in the phase diagram of the bulk FeSe$_{1-x}$S$_x$ is that the composition range where systematic single-crystalline samples are available is very small.
It is difficult to obtain samples with $x>0.2$ by the chemical vapor transport technique.
Although quite recently the growth of single crystalline samples with $x>0.2$ has been reported by a hydrothermal method\cite{PRB.96.121103}, the transport properties of these samples have not been measured sufficiently. 
Thin film growth of FeSe$_{1-x}$S$_x$ may also have an advantage that samples in a wider composition region are available. 
Recently the successful growth of FeSe$_{1-x}$S$_x$ epitaxial films with $x \leq 0.78$ was reported\cite{JJAP.56.100308}. 
However, these films did not show zero resistivity above 2 K for all $x$ values.

In this letter, we report the successful growth of superconducting FeSe$_{1-x}$S$_x$ films with $x \leq 0.43$.
As $x$ increases, $T_{\mathrm s}$ decreases, and $T_{\mathrm c}$ shows a gradual decrease even at the NEP.
In addition, samples with large $x$ values shows a rapid increase of the resistivity with a kink at low temperatures, indicative of a magnetic transition, which is not observed for bulk samples at ambient pressure.
In spite of the presence of the new phase transition, the behaviors of $T_{\mathrm c}$ and $T_{\mathrm s}$ are similar to those of bulk samples.
Considering the difference in the phase diagrams between FeSe$_{1-x}$S$_x$ and FeSe$_{1-y}$Te$_y$ films, most probable interpretation of our results is that the role of nematicity in superconductivity is not universal in iron chalcogenides. 

\begin{figure}
\includegraphics[width=\linewidth,bb=0 0 334 258]{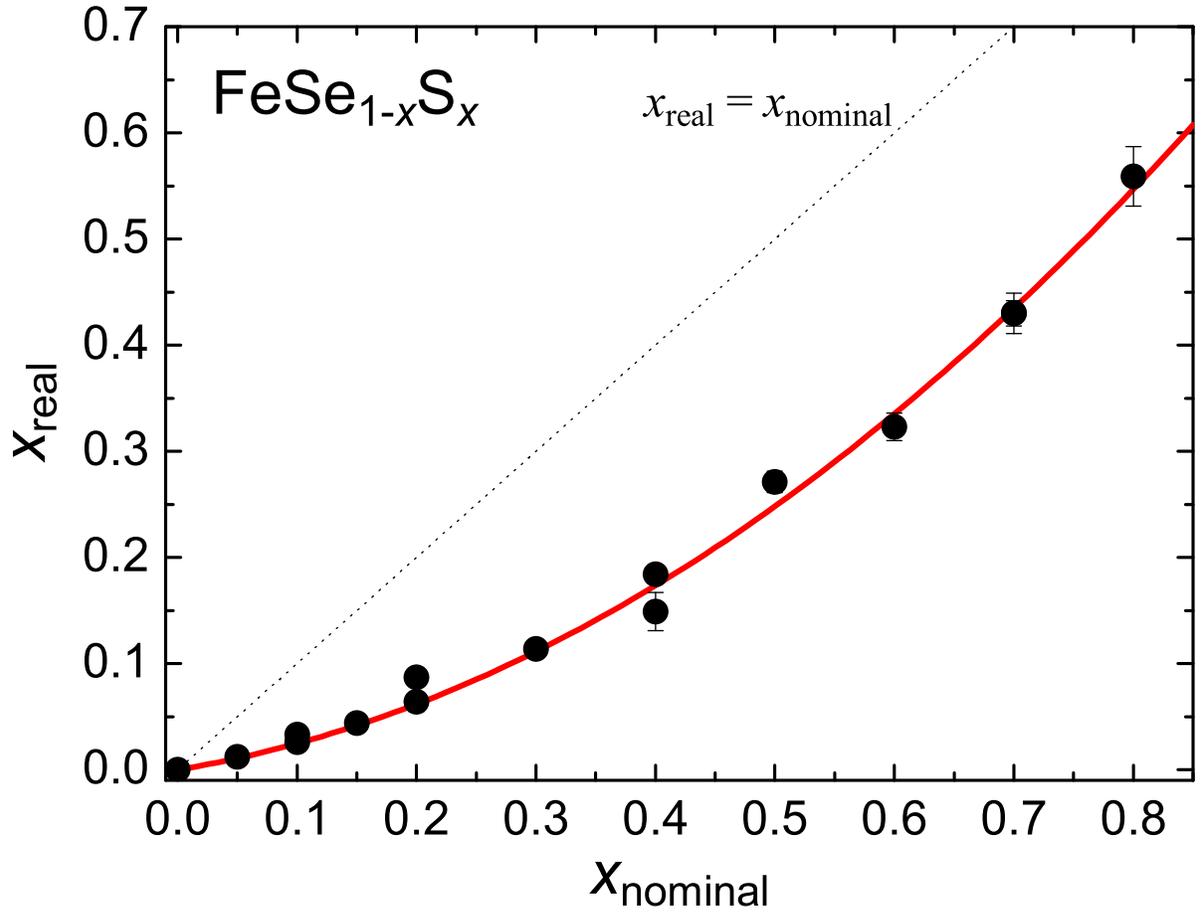}%
\caption{Real S content, $x_{\mathrm {real}}$, of the grown films, which was obtained by the EPMA measurements as a function of the nominal values, $x_{\mathrm {nominal}}$. Red curve shows the fitting result by a quadratic polynomial.}
\label{G_EPMA}
\end{figure}

All the FeSe$_{1-x}$S$_x$ thin films were grown by a pulsed laser deposition method using a KrF laser\cite{Imai09,Imai10}.
Commercially available LaAlO$_3$ (100) substrates were used in this study.
The substrate temperature, the laser repetition rate, and the back pressure were 300-400$^\circ$C, 1-3 Hz, and 10$^{-8}$ Torr, respectively.
FeSe and FeS polycrystalline pellets were used as target, which were alternatively ablated to obtain single phase FeSe$_{1-x}$S$_x$ films\cite{JJAP.56.100308}. 
Composition analysis was performed for some samples by electron probe micro-analysis (EPMA). %
Figure \ref{G_EPMA} plots the real composition of the films as a function of the $nominal$ values, i.e. the ratio of the ablation time for FeS to the total ablation time.
The real S content of the films increases up to $x_{\mathrm {real}}=0.43$ as $x_{\mathrm {nominal}}$ increases.
$x_{\mathrm {nominal}}$ dependence of $x_{\mathrm {real}}$ is well fitted by a quadratic polynominal, as shown by a red curve in the Fig. \ref{G_EPMA}.
Therefore we estimated the real composition of the other films with which a composition analysis was not performed, by using the empirical formula.  
Hereafter, the composition of the films, $x$, denotes the measured or calibrated values.
The crystal structures and the orientations of the films were characterized by four-circle X-ray diffraction (XRD) with Cu K$\alpha$ radiation at room temperature.
The thickness of the samples were evaluated by a Dektak 6 M stylus profiler. 
The electrical resistivity of the films was measured using a physical property measurement system (PPMS) from 2 to 300 K.

\begin{figure}
\includegraphics[width=\linewidth,bb=0 0 408 436]{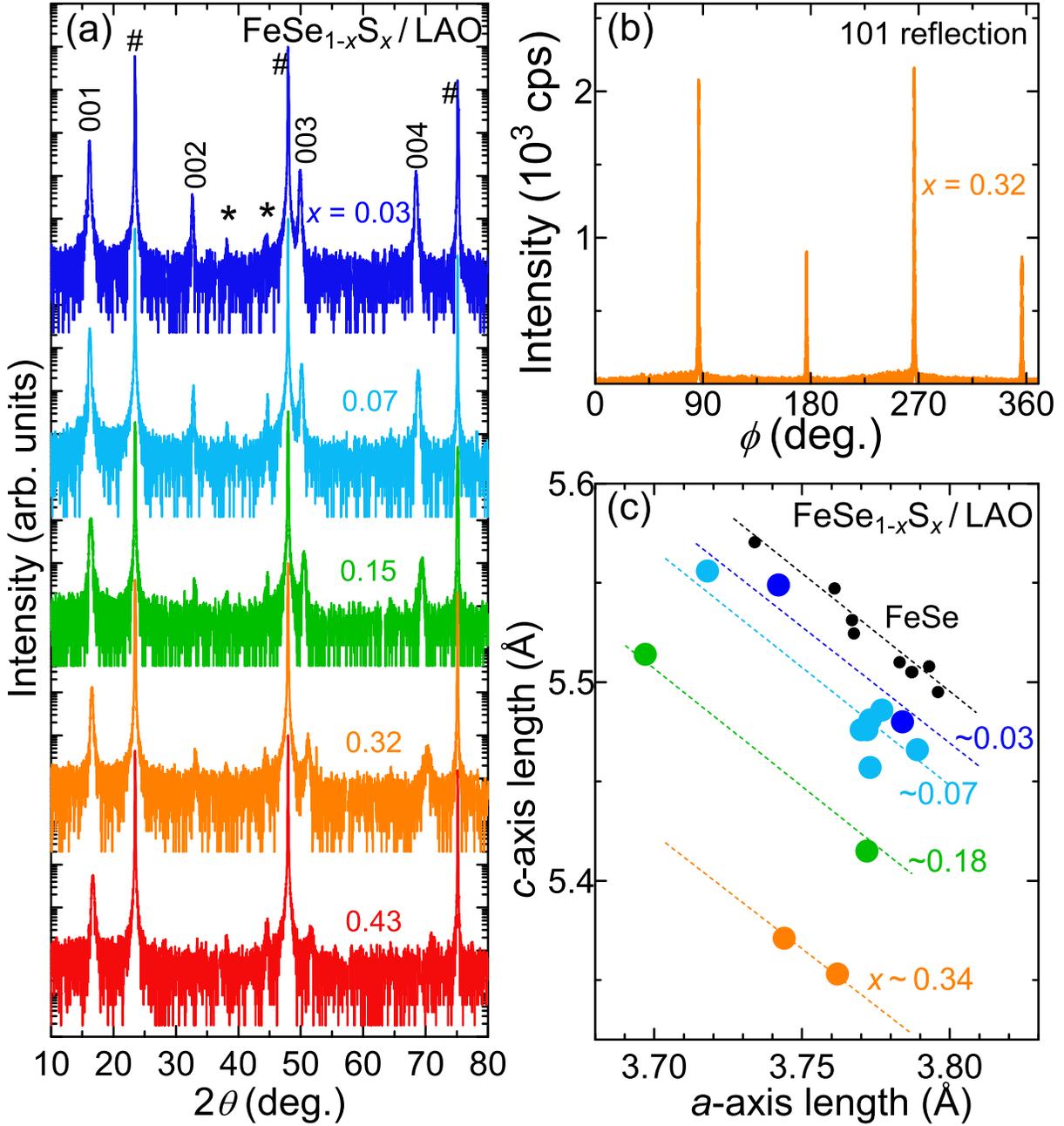}%
\caption{(a) XRD patterns of the grown films with $x=$0, 0.03, 0.07, 0.15. 0.32, and 0.43. The number and asterisk signs represent peaks from the substrates and the Ag paste, respectively. (b) $\phi$-scan of the 101 reflections for the film with $x=0.32$. (c) Relation between the $a$ and $c$ lattice parameters of the grown films. Data of FeSe films with varied degrees of strain are also plotted. The dashed curves are guides for the eye.}
\label{G_XRD}
\end{figure}

Figure \ref{G_XRD}(a) shows the XRD patterns of the films.
Except for peaks from the substrates, the buffer layers, and the Ag paste for the resistivity measurements, all the peaks are identified as $00l$ reflections of FeSe$_{1-x}$S$_x$, indicating the $c$-axis orientations of the films.
In-plane orientations of the films were also confirmed by the $\phi$-scans of the 101 reflections, which showed clear four-fold symmetry patterns (Fig. \ref{G_XRD}(b)).
We summarize the lattice constants of the grown films in Fig. \ref{G_XRD}(4), in which those of FeSe films are also plotted.
The lattice constants of the FeSe films change depending on the degrees of strain, and a negative correlation is observed between the $a$- and the $c$-axis lengths in the FeSe films, which is explained by the Poisson effect in films under the in-plane strain.
FeSe$_{1-x}$S$_x$ films with approximately the same composition shows the similar behavior, and the lattice constants tend to shrink as the $x$ values increase, indicating the validity of the calibration procedure of the $x$ values. 

\begin{figure}
\includegraphics[width=\linewidth,bb=0 0 435 439]{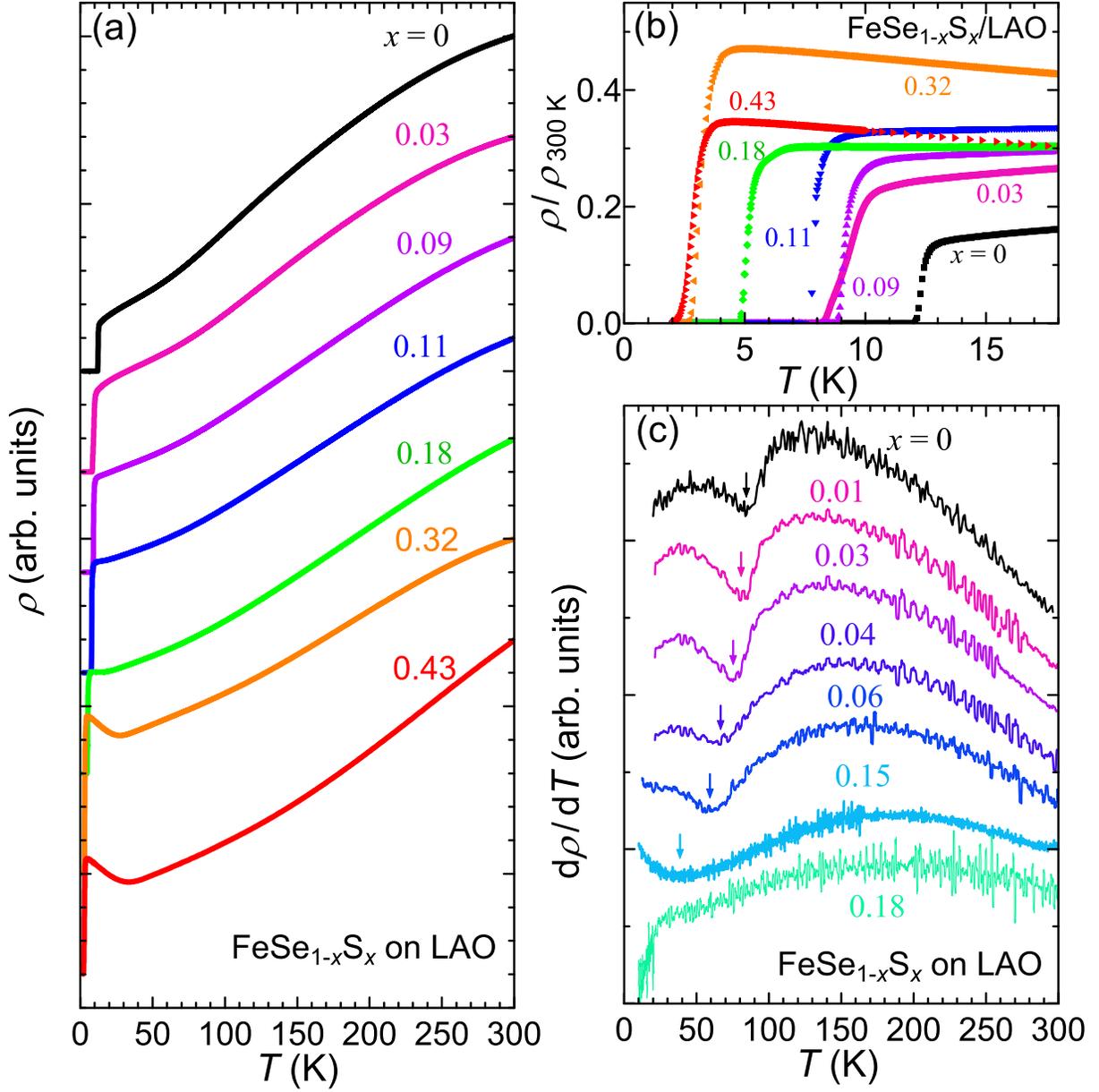}%
\caption{(a) Temperature dependence of the dc electrical resistivity, $\rho$, of the grown films. (b) Enlarged plots of the $\rho$-$T$ data around the superconducting transition. (c) Temperature derivative of the resistivity, d$\rho /$d$T$ of the grown films. The arrows show the temperature at which the anomalous dip in the d$\rho /$d$T$-$T$ curves was observed, corresponding to the nematic transition. }
\label{G_RT}
\end{figure}

Figure \ref{G_RT} shows the temperature dependence of the dc electrical resistivity, $\rho$, of the grown films.
All the FeSe$_{1-x}$S$_x$ films with $x \leq 0.43$ show metallic temperature dependences of $\rho$ and also show superconducting transition at low temperatures.
The FeSe film shows a kink anomaly in the $\rho$-$T$ curve, which is due to the nematic transition\cite{yiSciRep17}.
As S content increases, the kink anomaly becomes broadened gradually.
We estimated $T_{\mathrm s}$ of our films as the position of the local minimum in the d$\rho /$d$T$ curve, whose composition dependence will be discussed later.

FeSe$_{1-x}$S$_x$ films with large S content show another clear kink behavior at low temperatures, below which the resistivity increases on cooling.
The abrupt increase of the resistivity at low temperatures is very similar to what was observed in FeSe and FeSe$_{1-x}$S$_x$ bulk samples under hydrostatic pressure at an antiferromagnetic transition temperature\cite{JPSJ.84.063701,NatCom.8.1143}.
This similarity in the $\rho$-$T$ behaviors suggests that our films with $x>0.18$ also show antiferromagnetic order at low temperatures.
Note that in bulk samples this kink anomaly in the $\rho$-$T$ curve was not observed even in samples with $x=0.29$\cite{arXiv:1802.03776}.  
Thus this behavior is unique in film samples.
Although the reason for the appearance of the new phase transition only in film samples at ambient pressure is unclear at present, the lattice strain in the films may play a role.

\begin{figure}
\includegraphics[width=\linewidth,bb=12 339 584 818]{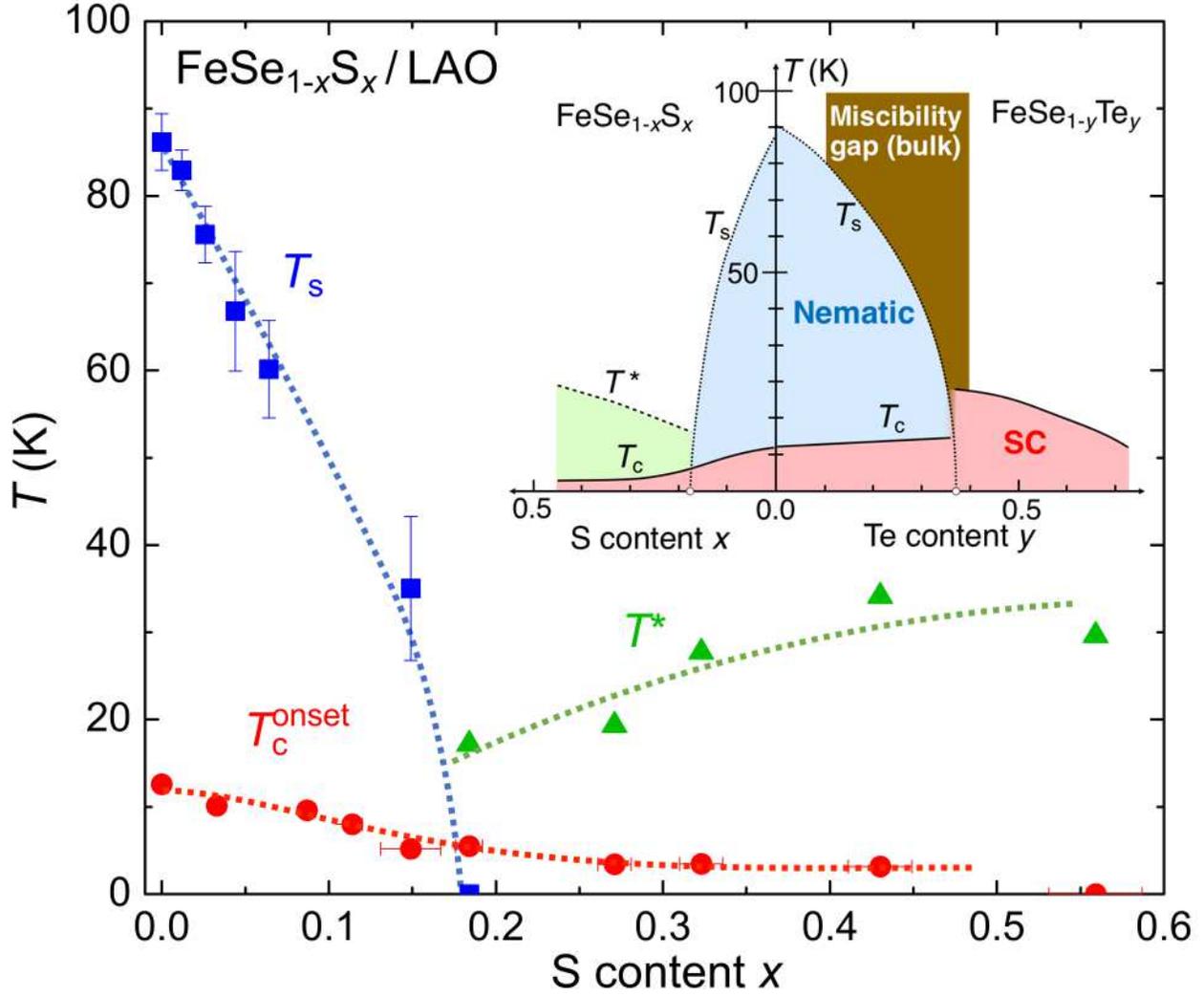}
\caption{Obtained phase diagrams of the FeSe$_{1-x}$S$_x$ thin films on LAO. The dashed curves are guides for the eye. The inset shows a schematic phase diagram of the FeSe$_{1-x}$S$_x$ and FeSe$_{1-y}$Te$_y$ films on LAO.}
\label{G_PhaseDiagram}
\end{figure}

Figure \ref{G_PhaseDiagram} shows the obtained phase diagram of FeSe$_{1-x}$S$_x$ films on LAO. 
The S substitution decreases $T_{\mathrm s}$, and the nematic order disappears at $x \sim 0.18$, similar to the case of bulk samples\cite{PRB.92.121108,PNAS.113.8139,arXiv:1608.01044}.
$T_{\mathrm c}$ also shows a monotonic decrease with increasing $x$ values. 
Note that bulk samples have the maximum value of $T_{\mathrm c}$ at $x\sim 0.07$, which was not observed for our samples. 
This slight difference in the behavior of $T_{\mathrm c}$ may be due to the variation in the strength of the lattice strain in our films.
With further increasing $x$, $T_{\mathrm c}$ gradually decreases even when $x$ exceeds the NEP, and no enhancement in $T_{\mathrm c}$ was observed.
We observed the superconducting transition for $x$ up to 0.43.
The behavior of $T_{\mathrm c}$ and $T_{\mathrm s}$ is very similar to those of bulk samples.
On the other hand, as was described above, what is different from bulk samples is the abrupt increase in $\rho$ at low temperatures, suggesting the presence of a new phase transition.
We plotted characteristic temperatures, $T^\ast$, at which $\rho$ takes its local minimum in Fig. \ref{G_PhaseDiagram}.
$T^\ast$ appears after the nematic state disappears, and increases with increasing $x$. 
There are two possible interpretations of the phase diagram.
One is that the new phase does not affect $T_{\mathrm c}$. 
The other is that the new phase suppress the superconductivity.
In the former case, we expect that $T_{\mathrm c}$ remains unchanged when the new phase is suppressed in some way, while in the latter case, we expect the increase of $T_{\mathrm c}$ at the NEP when the new phase is killed.
We consider that the latter case is unlikely because the composition dependence of $T_{\mathrm c}$ of the FeSe$_{1-x}$S$_x$ films is very similar to those of bulk crystals.
Thus the new phase observed below $T^\ast$ may not affect the superconductivity.
Further studies are needed in order to reveal the nature of this new phase transition and to elucidate the reason why it has no impact on superconductivity.

Once the phase diagram of FeSe$_{1-x}$S$_x$ films was obtained, we compare the phase diagram to that of FeSe$_{1-y}$Te$_y$ films.
As was described earlier, FeSe$_{1-y}$Te$_y$ shows the rapid enhancement in $T_{\mathrm c}$ when $x$ exceeds the NEP\cite{yiSciRep17}, which may suggests that the nematic order suppresses the superconductivity.
On the other hand, FeSe$_{1-x}$S$_x$ does not show such a $T_{\mathrm c}$ enhancement, or rather $T_{\mathrm c}$ decreases at the NEP.
These two contrasting phase diagrams demonstrate that the relation between the nematic order and the superconductivity is not universal in FeSe, suggesting that the nematicity does not have a universal significance in the superconductivity in these materials.
These results are consistent with recent results of the pressure effects in FeSe$_{1-x}$S$_x$ bulk crystals\cite{NatCom.8.1143}.
As was already clarified, the nematic transition will change the band structure and the Fermi surface\cite{PhysRevB.90.121111,PhysRevLett.113.237001,Nakajima17,Phan17}.
In that sense the nematicity will have some effect on the superconductivity in an indirect manner. 
Indeed, whether $T_{\mathrm c}$ increases or decreases after the nematic order disappears depends on the detailed structure of the Fermi surface of each material.  
Our results suggests that the nematicity does not play a primary role in the superconductivity in iron chalcogenides.
To understand the mechanisms of the two different behaviors of $T_{\mathrm c}$ in the phase diagrams of FeSe under positive and negative chemical pressure, comprehensive and systematic studies are indispensable using FeSe$_{1-x}$S$_x$ and FeSe$_{1-y}$Te$_y$ samples with a wide composition range.
These are uniquely possible with our film samples, which are now under way.

In conclusion, we have succeeded in growing superconducting FeSe$_{1-x}$S$_x$ thin films with $x \leq 0.43$ on LAO substrates.
As S content increases $T_{\mathrm s}$ decrease and the nematic state disappeared at $x \sim 0.18 $.
$T_{\mathrm c}$ gradually decreases with increasing $x$ and no abrupt increase in $T_{\mathrm c}$ was observed at the NEP.
The behaviors of $T_{\mathrm s}$ and $T_{\mathrm c}$ are very similar to those of bulk samples, while another phase transition was observed in film samples with large S content.
The difference in the behavior of $T_{\mathrm c}$ between FeSe$_{1-x}$S$_x$ and FeSe$_{1-y}$Te$_y$, which shows an abrupt enhancement in $T_{\mathrm c}$ after the disappearance of the nematic state, suggests that the nematicity does not play a primary role in the superconductivity of these materials.

\begin{acknowledgments}
We would like to thank K. Ueno at the University of Tokyo for the X-ray measurements.
We also thank S. Komiya at Central Research Institute of Electric Power Industry for the composition analysis of the films.
\end{acknowledgments}

\clearpage

\clearpage


\end{document}